# Beyond the Drawing Board: Toward More Effective Use of Whiteboard Content

Gene Golovchinsky, Scott Carter, Jacob Biehl

FX Palo Alto Laboratry, Inc.

3400 Hillview Ave., Bldg. 4

Palo Alto, CA 94304

gene,carter,biehl@fxpal.com

**ABSTRACT**

We developed a system that augments traditional office whiteboards with computation for the purposes of retrieving, reusing, and sharing whiteboard content. Our system automatically captures changes to whiteboard images, detects significant changes, and identifies potential collaborative activities. Users then browse and search the collection of images captured from their camera or shared from other users' cameras based on aspects such as location, time, collaboration, etc. We report on the results of a formative study and on an evaluation of effectiveness of our system, and discuss additional functionality that can be built on our framework.

**ACM Classification:** H5.2 [Information interfaces and presentation]: User Interfaces. - Graphical user interfaces.

**INTRODUCTION**

The use of whiteboards is pervasive across a wide range of work domains. Whiteboards enable users to quickly externalize an idea or concept, facilitate understanding among collaborators and peers, and can serve as conversational artifacts to ground discussion. Whiteboards provide a low-overhead work surface that allows information to be freely modified by both individuals and groups.

These affordances make whiteboards ideal tools to facilitate brainstorming and other associated creative activities such as reflection [1] and reinterpretation [1, 2]. Whiteboards also allow information content to be persistently visible after use, facilitating activity coordination and awareness of task progress, and supporting episodic memory.

To bring the affordances of whiteboards to digital tools, a wide range of commercial and research tools have been developed [3,4,5,7,15,16,17,18]. Despite their availability, however, use of these tools is all but nonexistent in the modern workplace while traditional whiteboards remain ubiquitous.

While there are many factors that contribute to this lack of adoption, we largely attribute it to the poor user experience provided by current electronic whiteboard tools. For instance, the overall fidelity of the digitizing hardware does not capture subtle nuances or expressiveness currently well supported by physical whiteboards. Further, many of these tools focus on the ongoing task, and offer poor support for retrieval and subsequent use of captured content. For instance, systems such as mimio (www.mimio.com) and eBeam (www.e-beam.com) that augment traditional whiteboard technology mitigate deficiencies of digital input, but provide little more than the ability to export portable images of content for post-task use; the burden of organization and management still falls on the user.

We believe the content creation affordances of traditional whiteboards to be essential to their overall success. In this paper, we present ReBoard, a new system that combines the affordances of existing whiteboards with complementary digital tools that facilitate the retrieval, repurpose, reflection, and use of whiteboard content long after its initial creation, whether or not it is still on the board.

ReBoard captures whiteboard content without explicit intervention from the user and stores content along with descriptive metadata. ReBoard provides multiple interfaces for users to retrieve previously-captured content by time, by position on the board, and by various other metadata, as well the ability to share content with peers. ReBoard improves captured content by correcting perspective distortion, by improving image contrast, and by compensating for changes due to light levels and for people moving through the field of view of the camera.

ReBoard is a novel augmentation of a traditional whiteboard that combines the natural interactions at a whiteboard to *create* content, with the power of a computer to *manage* it, thereby mitigating some of the limitations of both electronic boards and whiteboards. In the remainder of this paper we first explore existing work practices around whiteboards to establish some design requirements, then describe the architecture of ReBoard and its user interfaces, and then follow with an evaluation of the effective-ness of its capture algorithms. We conclude with a discussion of next steps.

**UNDERSTANDING WHITEBOARD USE**

Several studies have shown the value of whiteboards for externalizing ideas, for communicating difficult concepts to peers, for maintaining persistent access to important information, etc. For instance, Cherubini, *et al.* have shown that just within the domain of software development, whiteboards are used widely for design activities, for training, and for communicating with customers [1].

Teasley, et al.'s study of engineers at an automobile manufacturing company found whiteboards to be a preferred communication tool because they were always accessible and content created on the board could be easily viewed by all members of the collated team [2]. Studies within the domains of publishing [3], creative design [4], medicine [5] and research [6] have shown similar results.

Tang, et al. recently evaluated how the interaction characteristics of whiteboards impact their communication value and utility [7]. Their work reaffirms findings from *in situ* studies, showing whiteboards are effective at supporting both individual and collaborative work, serve as persistent displays of ongoing activities, and allow the purpose of the board, and the activities performed at them, to change fluidly. The study also highlighted limitations of whiteboards: Content could not be easily accessed by peers that weren't co-located; boards often filled quickly necessitating removing old, but still useful content; and no mechanism to understand how or why content changed over time.

Past research has tried to overcoming these limitations, while preserving affordances, by blending whiteboards with electronic capture. Systems such as Collaborage [8] and ZombieBoard [9] allow users to place marking artifacts directly on whiteboard content to direct the system to classify and route information to digital services. For example, users could update a status board and cause the changes to be processed and propagated onto the corporate intranet.

Several projects (e.g. [10,12,13,14,15,16,17,18]) and commercial projects (e.g. SmartBOARD (www.smarttech.com)) seek to mitigate the challenges by providing users a fully electronic whiteboards. For example, Flatland [10] provided users many of the affordances of traditional whiteboard, but also useful functionality for better managing content at time of creation. With the system users could, for instance, resize and move content on the board to make room for new markings or to organize ideas. Other tools, such as Tang and Minneman's VideoDraw [11] and Bly and Minneman's Commune [12] seek to extend the use of digital whiteboards to support distributed users.

While many of these tools do capture and preserve content created on whiteboards, their purpose is largely to provide support for an *ongoing* task. In contrast, ReBoard captures whiteboard content and lets users retrieve, manipulate, and share content *during* and *after* a task is completed. For instance, ReBoard users can retrieve content based on who was present during its creation, by a specific region on the whiteboard where it was created, by the time it was created, or by user-assigned labels and descriptions. This functionality allows users to leverage their episodic and spatial memories to recall content quickly.

EverNote (www.evernote.com) provides some capability for indexing and retrieving images, but does not automate capture in any way. Its indexing consists of time of capture, manually-entered metadata, and any text it can recognize in the image. However, considerable manual input is required to save each captured image. ReBoard, on the other hand, unobtrusively captures content passively and manages the archival and subsequent retrieval of content. This allows whiteboard users to focus on the ongoing task, without the distraction and overhead of having to explicitly capture and tag content. Additionally, ReBoard collects metadata appropriate to whiteboard content, rather than the generic data gathered by EverNote.

**Formative study**

While previous research has provided insights into the tasks supported by, and types of content created on whiteboards, much less is known about how content is later retrieved and reused. Since the latter was a central design goal of ReBoard, our design process started with building on the findings of previous work to create a richer understanding of whiteboard contents' use, including its creation, modification, sharing, and eventual archival or destruction.

We performed a small study with six research scientists recruited from within our organization. Largely following the methodology used in previous studies of whiteboard use (e.g. [8]) our study consisted of two phases: daily digital photographs of participants' actual whiteboard content (over the period of three weeks) followed by structured interviews. Content from whiteboards in public meeting spaces and a public "soft spaces" (e.g. a kitchen area) was also captured.

In the interviews, users were asked to explain how they used their whiteboard to facilitate their work activities and to describe how, why and by whom content was created and later used/reused. To aid elicitation of this information, we provided participants with printouts of images of their whiteboard (and any public whiteboards they used) and had them label and categorize the content into regions (e.g. logical groupings) based on the activity that content supported.

*Personal vs. Collaborative*

Across all participants, a total of 67 distinct drawings were defined ($\mu = 11.2$, $\sigma = 2.5$). Interestingly, we found use of the whiteboard was near evenly split between drawings supporting personal work (37/67, ~55%) and collaborative drawings (30/67). We also analyzed the total area covered by different types of content. Personal drawings covered

45% (σ = 7%) of the boards, collaborative drawings 34% (σ = 10%), objects attached to the board (such as paper fliers) 5% (σ = 2%), and the remainder was either whitespace, errant strokes, or old drawings whose meaning participants could no longer recollect.

Drawing type closely matched the physical utilization of a whiteboard's marking surface: 57% of the marking area was dedicated to personal drawings, while 43% was collaborative. We also found that the location of drawings was somewhat related to the type of activity they supported. Collaborative regions were more often located on the half of the board closest to the entrance of a participants' office (63% (σ = 14%)). Likewise, personal content was more often created on parts of the whiteboard farthest from the office entrance (58% (σ = 5%)). Although not unexpected given the geometry of our offices, these results do suggest that there was some loose organization based on the associated activity.

### Recall and Reuse

Over the course of the study, we found that, on average, 84% (σ = 15%) of participants' whiteboards marking area contained content. However, only 52% (σ = 14%) of the marking area changed from the beginning of the study to the end. Interviews with participants revealed three interdependent reasons for why some content remained static. First, participants indicated they typically only erase content when space for new content is needed. For instance, many participants reported that their whiteboard was often full when they wanted to use it, and it was only then that they put thought into deciding which region to remove.

Second, participants indicated a significant amount of content was persistent. Such content consisted of to-do items, phone lists, web site addresses, etc. Finally participants also indicated that they felt that even ephemeral content was useful as a reminder and/or reference of ongoing or planned activities. For instance, a design schematic of a software API could be repeatedly consulted to recall design details.

### Limited Reuse

Interestingly, while participants indicated whiteboard content had value well beyond its creation, very few participants indicated they took proactive measures to protect or archive the content. For example, most participants found it too time consuming to take a digital photo of the whiteboard. One of the six participants did indicate he would occasionally take a digital photo, but would often forget where he had stored it at the time of desired reuse.

In our interviews, we also found that users would use whiteboards in sporadic bursts. It was common, for example, for a collaborative session to generate multiple erasures and markings on the board. Afterwards, only the final markings could serve as persistent content. All participants indicated this to be a major drawback of using whiteboards as the final markings were often not sufficient to recall or recreate ideas or concepts that were overwritten.

### Key Design Challenges

Leveraging the insights gained through our formative study along with lessons provided by past work (e.g. [8,9]), we derived the following key challenges for the design of systems that facilitate effective use of whiteboard content:

- There is high perceived value in electronically capturing whiteboard content, but the overhead of existing tools and methods is a barrier to the practice.
- Whiteboards are successful collaborative tools at the time of content creation, but their physical characteristics make it difficult to distribute and share content.
- The purpose and type of activity often dictates the spatial location of content created on the whiteboard. The location is often used to assist in recall and reuse content. For example, phone numbers and to-do lists are often found at the edges, while more transient diagrams, notes, etc. tend to be written in the center.
- Regions of content on whiteboards are often used as peripheral displays, providing a user an always-on reminder of critical task-related information.
- Often during whiteboard activities a significant amount of whiteboard content is rapidly created and destroyed. This presents methodological and social constraints for capturing in-task content.

In the next section we describe how we addressed these challenges in the design of ReBoard, a system that facilitates not only lightweight capture of whiteboard content, but also provides a set of novel tools for managing the retrieval and reuse of captured images.

### REBOARD

The ReBoard system is designed to capture whiteboard images unobtrusively, and to facilitate their subsequent reuse. The following scenario illustrates how ReBoard enables more effective whiteboard use:

*Alex and Ben are software developers who are meeting to discuss the design of a new module. They take turns sketching out UML and other related diagrams on Ben's whiteboard to design the mechanics of the new module. Ben uses ReBoard's persistent UI loaded on a Chumby located next to his whiteboard to capture intermediate high-level drawings, before they dive into the detailed UML diagram.*

*Alex and Ben become engrossed in the task and lose track of time. Suddenly, Alex remembers that the staff meeting that is about to start. They immediately leave Ben's office, but forget to capture the last diagram created. Fortunately, ReBoard automatically detects the changes and captures an image of the content.*

*After the meeting, Ben discusses the module design with Charlie and he becomes intrigued with some of the details. Ben returns to his office and uses ReBoard to share the captured diagrams with Charlie. Ben also identifies Alex as a contributor to the content, thus allowing both Charlie and Alex to retrieve the diagrams.*

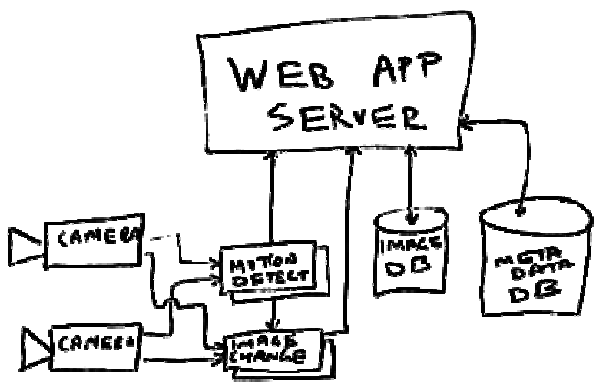

Figure 1: ReBoard architecture (from a ReBoard image).

*A month later, Alex needs to prepare for an upcoming design review. Because Alex does not know the specific date the diagrams were created, he first uses ReBoard's calendar view to select the approximate date, and then switches to the timeline view to quickly scan clusters of captured content. With little effort Alex finds the diagrams and pastes them into his presentation notes.*

While simple, this scenario illustrates many of the novel capabilities of ReBoard. These include the ability to both actively and passively capture whiteboard content, to share access to captured content and specify contributors, and to retrieve content using a variety of metadata the user might happen to remember.

The ReBoard system architecture (Figure 1) consists of a battery of video cameras (one per whiteboard), an application server, a database to store archived images and associated metadata, detectors that sense collaborative activity at boards, and detectors that record whiteboard content changes, and user interfaces to manage content and metadata. The following sections describe each of the main components in more detail.

**Application server**
The web application server is the heart of the ReBoard System: It manages the data collection and retrieval of images, and coordinates the actions of the event detectors. The server is written in Grails[1] and stores its metadata in a MySQL database. The server manages user accounts, sessions, access control, and exposes a set of REST-like web services for AJAX UI clients and for communicating with the event detectors. In addition, an asynchronous messaging system is used to communicate event notification between collaboration and content change detectors.

The application server actively manages how event detectors are allocated to the whiteboard cameras. To facilitate this, each event detector is assigned a unique id; when the event detector process comes online, it queries the application server with its id to receive its camera assignments. Event detectors periodically poll the server for updates to these assignments, allowing us to adjust the mapping in near real-time without restarting services.

This architecture allows the detectors to scale and operate independently. In our prototype, we have deployed the application server and a cluster of event detection services on our corporate cloud computing infrastructure. The combined software and hardware scalability allow easy reconfiguration of system components to respond to changes in demand and/or deployment characteristics. It also allows experimental components to co-exist with production components, improving our ability to iteratively prototype changes and new functionality.

**Detecting collaborative activity**
A novel feature of ReBoard is the ability to search and browse captured whiteboard content by specifying whether or not the content was created as part of a collaborative activity (i.e., more than one person being present). The process of automatically tagging content as collaborative is performed as a two-step process: detecting collaborative activity at the whiteboard, and then labeling the associated captured whiteboard content as collaborative. It is important to note that this detector does not do person tracking *per se*; rather the blobs are taken to represent people simply because it is unlikely that anything else will move between the camera and the board. This assumption simplifies some of the signal processing required to detect potential collaborative activity.

To detect collaborative activity, we leverage the existing 10fps video stream used for content capture to also monitor the physical space in front the whiteboard for motion. The motion event detector continuously performs simple background differencing between frames to detect changes. This low-pass filter greatly reduces triggering false positives from anomalies in the camera signal (e.g. compression artifacts) and small amounts of apparent motion in the physical space (e.g. small changes in lighting and/or reflection off the whiteboard surface).

If more than 5% of the all pixels in a frame change, we then apply a Gaussian Mixture Model to cluster pixel-level motion into rectilinear motion blobs. Overlapping blobs are also combined into single, larger blobs, and blobs that are either too small or too big to be people are excluded. The resulting blobs are counted to approximate the number of people currently in front of the whiteboard.

At this point, the motion detector notifies the content change detector that motion has occurred near the board. It then further filters the motion data to determine the beginning and end of collaborative events.

For each camera, the event detector keeps a five minute history of the number of people present in each captured frame. Every 15 seconds, it calculates the mean number of people in the most recent 2.5 minutes of the detection history. If the average is above an environment specific threshold (around 1.8 for most camera installations), it sets the current time as the start of a collaborative activity. The system then transitions to calculating the mean over the entire five minute history. When the value falls below another environment specific threshold (1.3 in most installations) the current time is specified as the end of activity and the event is sent to the application server.

---

[1] Grails is a web application framework built on the Java VM. See www.grails.org

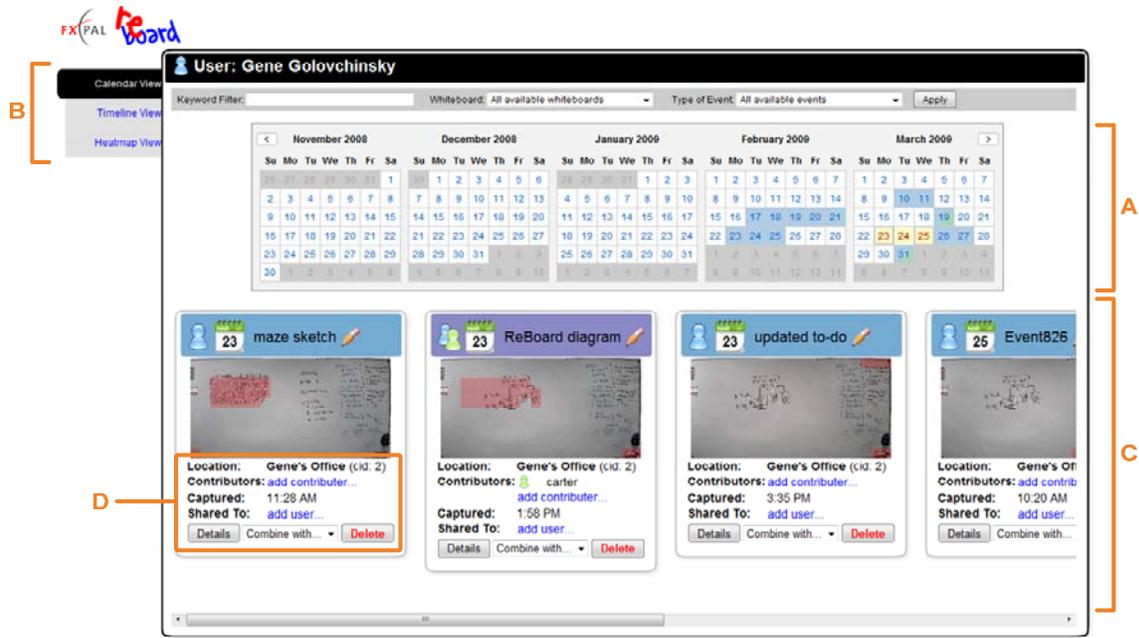

Figure 2: ReBoard's web UI, the main interface for browsing and retrieving captured whiteboard content. The interface has two main components: a pane for providing overview visualizations of captured content and a pane for previewing details of selected content. In this screenshot, the *calendar* view (A), which summarizes captured content using a color-coded array of calendar widgets, was selected from the list of available views (B). Three dates (March 23-25) in the calendar view were selected which generated previews of content occurring on those dates (C). From each preview (D) a user can add (or remove) contributors, share content, open a detailed view (see Figure 3), etc.

Associating a detected collaborative activity with corresponding whiteboard content is performed as events are received by the application server. When a collaborative activity event is received, all images that were also captured on the same camera within the collaboration activity interval are grouped and labeled as collaborative. Likewise, as image capture events are received, the server records those images as collaborative if they fall within the intervals of already recorded collaborative activities. This approach allows manually-recorded images and automatically-captured images to be integrated gracefully with automated collaboration detection; the result is a smoothly-running mixed-initiative system.

In the current prototype, this detector is a C# Windows Service that uses the AForge (code.google.com/p/aforge/) computer vision framework for basic video processing and blob detection.

**Detecting content changes**

ReBoard can support a number of approaches for capturing whiteboard content. For example, ReBoard can accept stroke events from whiteboard augmentation tools that rely on ultrasonic sensors attached to pens (such as eBeam). However, systems that require special devices have many of the same limitations as electronic whiteboards, most importantly that they cannot capture all whiteboard content, such as objects attached to the board. For this reason, our current implementation relies on camera images.

ReBoard's whiteboard content change detector is designed to ignore changes due to people or other objects moving through the field of view of the camera, and to changes due to varying light levels. Cameras are calibrated at installation to correct for geometric distortion due to off-axis camera position. Specifying the coordinates of the whiteboard's four corners and its aspect ratio, we can compute the transform required to correct the distortion. We then crop non-whiteboard portions of the camera image to avoid detecting changes in the environment (rather than changes on the whiteboard). Cropping also makes the images much more readable to users.

To determine if the whiteboard is in a state suitable for image capture, the content change detector first checks if the motion detector (described in the previous section) detected motion since the last update. If not, the content update detector assumes that any changes to the board must be due to noise and does not record any updates. However, if there has been motion at the board since the last update, the detector attempts to capture an unobstructed image of the whiteboard. To do this, it grabs several images from the camera at 200ms intervals and computes a pixel difference between the images. If significant differences are found among the images (calibrated for environmental conditions), the detector assumes a person is moving through the camera's field of view and discards these images.

The detector also analyzes the brightness of these images – if it is too low, it is likely that no one is working in the office and the images are discarded. The detector also looks for large static regions (such as chairs) that might be blocking the board. If these are detected, the image is not used.

Once the detector records an unobstructed image, it applies filters to ensure the cleanest possible images for analyzing content changes. To accentuate strokes while minimizing lighting irregularities, we apply a high pass filter, using kernel $\{0,-1,0;-1,k,-1;0-1,0\}$, where $k$ varies per camera. This removes much of the low frequency glare and shadow noise from the image. A low pass filter using kernel $\{1/9,1/9,1/9;1/9,1/9,1/9;1/9,1/9,1/9\}$ is then applied to eliminate high frequency camera noise.

After this filtering, a pixel difference is computed against the image previously determined to be *actual* changed whiteboard content. If the pixel difference is above a camera specific threshold, the image is the considered to contain new changes in the whiteboard's content. For these images, the detector performs further processing to compute per-region pixel differences for the image. Two sets of square regions are computed: an *X*-by-10 low-resolution set and a *X*-by-100 set. *X* is determined by the aspect ratio of the board (e.g. a board with aspect ratio of 1.6 would have a 16-by-10 low-resolution set). This information is used by ReBoard's user interface to generate change highlights and the *heatmap view* (see next section).

The image, pixel change region sets, and other metadata (e.g. location and time of capture) are sent to ReBoard's application server for archiving and later retrieval.

In the current prototype, we implemented the content change detector as a Java-based service that leverages Java's Media framework and OpenCV's image processing libraries.

**User Interface**
While our system relies on the traditional whiteboard for content creation, electronic interfaces are provided for retrieval, reuse, and sharing of captured content. ReBoard's architecture can enable a large variety of custom interfaces. In this section we present two interfaces that are currently implemented: a web-based interface and lightweight persistent interface. We discuss these in turn below.

*Web UI*
ReBoard's web interface was designed to be the primary means of retrieving and reusing images captured from instrumented whiteboards. The interface allows users to pivot quickly across three different, yet complimentary, views: *calendar*, *timeline*, and *heatmap*.

The *calendar view* summarizes the captured content using an array of interactive calendar widgets (see top of Figure 2). Dates on which content was captured are highlighted in a color corresponding to the type of captured content. For example, blue highlights represent personal content, violet for collaborative, and green for shared. If more than one type of event occurs on a particular date, the date is striped with the appropriate colors.

Users can select a single date or range of dates to show summaries of content captured in that time span. Selected days are highlighted in yellow and a *summary pane* of the captured content on the selected date(s) is displayed below the calendar array. The summary allows users to browse thumbnails of the captured images and the associated metadata; e.g. when the image was captured, with whom it was shared and, whether it was created collaboratively. The view also (optionally) highlights the changed regions as a partially transparent overlay on the thumbnail images.

The *timeline view* (Figure 4) arranges captured images chronologically. Each captured image is represented by a bar in a histogram: the taller the bar, the more content was created or erased for that capture. Dates at the bottom of the timeline assist approximate navigation. Complementary to the *calendar view*, this view is useful when the user knows only the approximate date of capture.

Users can drag a highlighted region on the timeline to change the captured content in the *summary pane* to correspond to the highlighted span of time. This provides a lightweight mechanism quickly to browse captured images over a fairly large span of time.

The *heatmap view* aggregates changes to the board by location over a given range of time. This view enables users better to leverage their spatial memory of *where* content was created or modified to assist in retrieval.

In this view, the more a region of the board was changed, the warmer the color that is used to fill the region. For example, a region of the board that was changed many times over a given time span is colored in red, while a region that only changed once or twice is colored in blue. Regions that never change are colored in white. Regions of the heatmap can be selected (via a click, drag, release mouse interaction) to populate the *summary pane* with captured content that had been added or erased in that region. Calendar widgets are used to define the time range over which content will be aggregated.

In the *summary pane*, users can click the detail button for a particular content record to open new browser tab (Figure 3), that provides a detailed interface for viewing high-resolution images of the captured content and controls to manipulate associated metadata (e.g. tags and description). The content owner (e.g. the person in whose office the content was captured), can set identity of contributors and specify with whom to share the content. Users can choose to share the entire whiteboard or just a particular region. While it can be overridden, the default is to share only the bounding rectangle of the largest region of detected change. This protects potentially private or irrelevant information on other regions of the whiteboard.

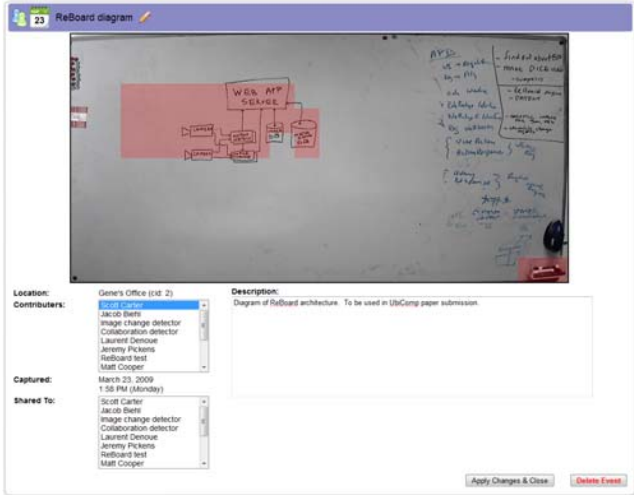

Figure 3: A detailed view of captured content. Users can manipulate associated metadata; including who contributed to the creation of the content, with whom it is shared, and its label and description.

An important aspect of ReBoard's UI design is that it preserves retrieval parameters *across* views. Thus, for a single retrieval task, users can leverage multiple views to find the target content. For example, a user can first use the *heatmap view* to select a region of the whiteboard, and then switch to the *timeline view* to visualize content captured *only* within the region selected in the *heatmap view*.

In addition to leveraging the different views, users can restrict retrieval content based on other metadata filters. These include capture source (location of the whiteboard), type of content (personal, shared, or collaborative), and keywords (on contents' label and description text). In the current prototype, these filters can be set in any view, using dropdown menus and textboxes at the top of the view.

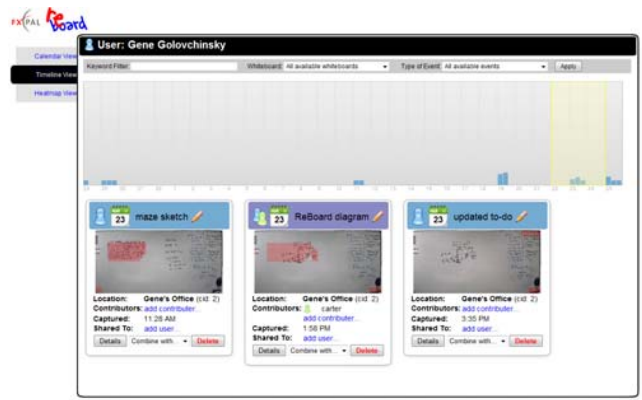

Figure 4: ReBoard's web UI showing the *timeline* overview. The height of each bar in the timeline corresponds to the number of regions that changed at the time of capture.

### Persistent Peripheral UI

Supporting findings from our formative evaluation, as well as from prior research, we provide a persistent UI to support *within-activity* explicit control of the system actions. Such control could be used to manually initiate a capture, or to disable automatic capture of sensitive or private content.

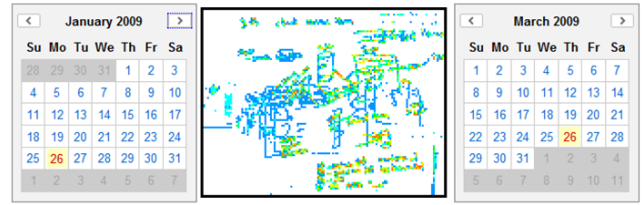

Figure 5: A close-up of the *heatmap* view showing two months of whiteboard content. Note how salient features of the content "show through" enabling easy spotting of familiar content.

We felt using the web-based UI to perform this level of control would be too burdensome to users. For instance, a user would have to stop her activity at the whiteboard, physically move to the desktop or laptop computer, load the ReBoard Web UI, perform the appropriate actions, and then physically move back to the whiteboard. Thus, we developed a persistent UI that runs on a small portable device placed nearby to provide immediate, low overhead within-activity capture control to capture images manually, to disable capture, to bookmark specific content, and to preview current and recent captures.

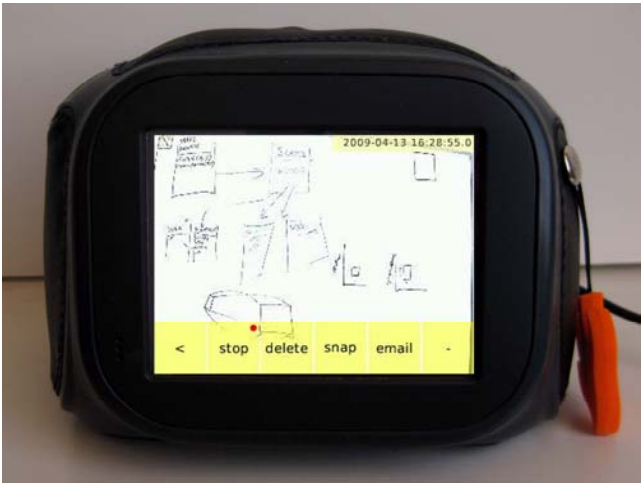

Figure 6: The ReBoard persistent user interface running on a Chumby device. The interface provides one-touch manual capture and a persistent display of content archived by the ReBoard system.

We implemented our persistent UI prototype using the Chumby device/application platform (www.chumby.com), shown in Figure 6. While there are many alternative devices, the Chumby was chosen because it has a bright and responsive touchscreen, an aesthetic exterior, and ability to be remotely configured. It is important to note that the persistent peripheral UI, while useful, is not necessary for deployment of ReBoard-enabled whiteboards.

**EVALUATION**

To gauge the effectiveness of our system, we compared the results of our image capture and collaboration detection algorithms against hand-coded data. We also collected data for two other image capture approaches: one using image filtering but not motion detection, and one using motion detection but not filtering.

We collected data for all three approaches during a pilot deployment of the system over 19 days for 10 different participants. We furthermore hand-coded video archived over this period (video was logged from each camera at 1 frame/second at VGA resolution), noting the time and date of each whiteboard update as well as whether the change involved collaborators. In our analysis of the three algorithms, we considered a whiteboard content change valid if it was recorded within a 15 minute window on either side of the ground truth event.

While all participants spent at least part of the study period working in their office, their schedules varied widely: many were away for several days while others worked through nights and weekends. In order to be sure that we compensated for each participant's schedule, we ran all capture algorithms continuously. However, we did redact 4 days worth of data from 7 cameras that had technical difficulties.

For both the complete and filtering-only algorithms, we determined the high pass kernel value $k$ and pixel threshold $t$ a priori such that our algorithm would detect a mark covering a region $1/100^{th}$ the area of the display. For the complete and motion-only algorithms, we logged a motion event if 5% of the image changed over 1 second.

Our results showed that motion data plays a large role in avoiding erroneous whiteboard updates. Hand-coding revealed a total of 81 whiteboard updates. The combined algorithm recorded 69% (32% RSD) of these changes while capturing only 1.05 (1.11 SD) false positives per day per camera on average. Similarly, the motion-only algorithm found 64% (28% RSD) of captures, while logging only 1.17 (.98 SD) false positives. However, while the filtering-only approach found a similar amount of updates – 60% (23% RSD) – it captured 6.97 (2.11 SD) false positives per day per camera.

Importantly, a majority of the whiteboard update events that our combined algorithm did not capture were either small, incremental updates or situations in which some object (usually an office chair) obstructed the board for a long period of time after the update. In both cases, the algorithm would log the update eventually, either after several incremental updates or after the obstructing object was moved from the board. In fact, we found that the combined algorithm logged 88% of whiteboard changes (31% RSD) at some time, and those that were not captured tended to be small and quickly erased. In summary, ReBoard captured most whiteboard updates while keeping false positives to a minimum.

Lastly, while hand-coding video streams we noticed that during collaborative sessions office visitors would often spend the bulk of their time around the threshold of the office, occasionally pick up a pen, edit the board, and then return to their perch next to the door. While this is interpretable to a human as a collaborative event, it is problematic for a camera that is trained on the whiteboard. To address this issue, we are in the process of installing in participants' offices a secondary, lower-resolution camera with a wider field of view that can better capture collaborative activity anywhere in the office.

**DISCUSSION**

There are several important aspects to the work described in this paper. We have demonstrated a novel system that combines low-overhead capture of whiteboard images with multi-faceted retrieval interfaces that leverage people's episodic memory to help recover previously-created content. Unlike other whiteboard capture systems, ReBoard uses rich spatial, temporal and collaborative metadata to facilitate retrieval and subsequent reuse of captured images.

The approach is not without limitations, particularly compared with respect to digital whiteboards. Digital boards that capture marks as strokes are better at manipulating these marks after the fact, and may more readily apply handwriting recognition algorithms to the stroke data. In addition, the stochastic recognition processes employed by

ReBoard will never be 100% accurate, requiring the users to occasionally delete extraneous captures. We do not believe, however, that these limitations are fatal: the lack of penetration of digital whiteboards into the everyday office environment (compared with conventional boards) suggests that whiteboards are more effective content creation and collaboration tools; the introduction of capture tools such as ReBoard should only increase their overall usability.

The early pilot deployment of ReBoard has generated some insights into its utility that we were not expecting. For example, users have noted that they erase their boards more frequently, confident that the content would be saved by ReBoard. These users have begun to rely on ReBoard to capture drawings and other whiteboard content, and have regularly referred to (now erased) content.

In meetings, these pilot users also commented that they take fewer notes in paper notebooks, and instead write on whiteboards and share the captured content. The advantage of this approach lies in part in consolidating notes in one place, making it possible to retrieve them for future reference. ReBoard has also shown useful for supporting asynchronous self-collaboration, including transitions from synchronous to asynchronous work as described by Tang [9].

One of the most interesting pieces of feedback from the pilot deployment is that naive overlap of content images in the *heatmap view* has proven surprisingly useful, as familiar diagrams and notes appear to "shine through" other ink (Figure 5). Although more experimentation is required, the spatial view of captured content provided by ReBoard's *heatmap view* may provide a useful mechanism to navigate whiteboard history. These overlapped images may also serve as compact reminders of recent activity.

Automated systems are rarely perfect at inferring users' intent. Similar to past work that utilized mixed initiative approaches to resolve ambiguities in sensor data [19] and resource allocation [20], our system allows users to initiate capture when they want, but backs that up with automated capture based on heuristics that characterize likely important changes to whiteboard content. The nature of change detection suggests that false-positive errors are more common than false-negative ones, allowing the user to remove unneeded extra images that are captured occasionally.

## FUTURE WORK

ReBoard can serve as a platform for exploring several aspects of whiteboard use and capture. We discuss these directions below.

### Study

In this paper we describe the process we followed to design and prototype a new system for improving utility of whiteboard content. An important next step of our work will be to study how the ReBoard system is used and to utilize this understanding to further refine its design and functionality. We believe the best way to accomplish this is through a longitudinal field deployment across a variety of users. This will provide not only insights into ReBoard's impact on whiteboard use, but also the impact on overall communication and collaboration behaviors.

### Reminders

As we found in our literature survey and formative work, people often use parts of whiteboards for persistent visual reminders. Over time, these reminders can consume a large part of the whiteboard, making it unavailable for other purposes. While erasing these reminders can free up space, the value of the reminders is lost.

We are currently developing a method of extracting reminders from images of people's whiteboards, and displaying these reminders as overlays on the desktop or in a user's personal information management (PIM) software (e.g., MS Outlook). Once captured, the reminders can be left on the whiteboard, or erased. The user will be able to retain the reminder aspects online, while freeing up the whiteboard for other uses. Furthermore, reminders can be created retroactively from previously-saved whiteboard images.

### Handwriting recognition

It may be useful to recognize some of the handwriting that people use on the board to help with retrieval when the topic of the discussion is known. Although the signal may be quite noisy due in part to lack of temporal stroke information, it is possible that some terms may be extracted. These should be added to the tag and label index already in place to improve the chances of finding the right image. Alternatively, integration with eBeam or other stroke capture technology (that still uses traditional dry-erase markers) may provide an additional channel for capturing stroke data.

## CONCLUSION

Drawing on the available research literature and on our own observations, we designed set of augmentations for existing whiteboards to support existing practices around creation of drawings and notes, and to leverage the power of the computer to manage that information once created.

We built a system that captures whiteboard images unobtrusively, without disrupting existing work practices around content creation. The system's user interfaces allow users to browse and search through the captured images to retrieve already-erased content. The system also allows images to be shared with other users, so content created in one location can be accessed by different people in different places. The system identifies images based on associated metadata, including the time and location on the board when the image was captured, and whether it was created collaboratively.

We evaluated several techniques for image processing to improve the accuracy of automatic image capture, and found that a combination of image- and motion-based cues resulted in the most accurate change detection. Overall, this suggests that a mixed-initiative approach that combines reasonably reliable automated capture with manual intervention for capture at specific points and for occasional error correction is viable in this design space.

Our users have embraced the system, and have been using it during the initial deployment. We are looking forward to a broader roll-out of this technology in a couple of months.

**ACKNOWLEDGMENTS**

We thank John Adcock for the image de-skewing code, FXPAL management for supporting this research, and our colleagues for their helpful comments. We thank Art Muir for his help with the eBeam SDK.